\documentclass{epl}
\usepackage{amsmath}

\providecommand*{\dd}{\mathrm{d}}
\providecommand*{\dod}[2]{\frac{\dd #1}{\dd #2}}
\providecommand*{\op}[1]{\hat{#1}}

\providecommand*{\Hop}{\op H}
\providecommand*{\Tr}{\operatorname{Tr}}
\providecommand*{\kom}[2]{[#1,#2]}
\providecommand*{\Ket}[1]{\left|#1\right>}
\providecommand*{\Bra}[1]{\left<#1\right|}

\title{Small quantum networks operating as quantum thermodynamic machines}
\author{M. J. Henrich\inst{1}\thanks{E-mail: \email{Markus.Henrich@itp1.uni-stuttgart.de}} \and M. Michel\inst{1} \and G. Mahler\inst{1}}

\institute{                    
  \inst{1} Institute of Theoretical Physics I - University of Stuttgart,
Pfaffenwaldring 57, 70550 Stuttgart, Germany
}

\begin{document}

\maketitle

\begin{abstract}
We show that a 3-qubit system as studied for quantum information purposes can alternatively be used as a thermodynamic machine when externally driven in finite time and interfaced between two split baths. The spins are arranged in a chain where the working spin in the middle exercises Carnot cycles the area of which defines the exchanged work. The cycle orientation (sign of the exchanged work) flips as the difference of bath temperatures goes through a critical value.
\end{abstract}

Back in 1824 Sadi Carnot published his "Reflections on the Motive Power of Fire", proposing ideas for a combustion engine \cite{hyperjeff}. His reasoning was based on his now famous ideal gas cycle analysis. With the advent of nano physics the intriguing question arises: Down to which scales can one reasonable define thermodynamic variables and thermodynamic machines?

First attempts on the quantum level have been made in \cite{Scovil1959, Geusic1967}. Meanwhile such miniature machines have attracted quite some attention \cite{Feldmann2003, Palao2001, Segal2005, Bender2000, Allahverdyan2005, Kieu2006}. Artificial autonomous nanomotors powered by visible light have recently been demonstrated experimentally \cite{Balzani2006}. Nanomechanical devices have been of growing interest in the last couple of years \cite{Schwab2005}; the importance of cooling processes is being recognized with respect to quantum information processing \cite{Baugh2005}. Furthermore, it has been realized that strange analogies appear to exist between the basic concepts of quantum information theory like specific entanglement measures on the one hand side and thermodynamic concepts on the other side \cite{Oppenheim2002}, the origin of which is still fairly unclear.

Two level systems (TLS) like spins or qubits are the essential ingredient for quantum computation. Much effort has been directed towards control of small clusters and chains of qubits in quantum optical systems \cite{Cirac2000}, nuclear magnetic resonance \cite{Gershenfeld1997} and solid state systems \cite{Makhlin1999}. A serious problem in any such realisation is the interaction of the respective quantum network with its environment.

Here we argue that structurally identical quantum networks can be used as quantum information processors as well as quantum thermodynamic engines! This may not only shed new light on that intricate relationship, but may also give new directions for the application of such few-TLS structures, for which a number of different implementations are becoming available now.

\begin{figure}
\centering
\includegraphics[width=.45 \textwidth]{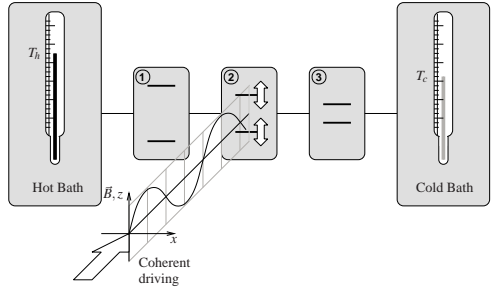}
\caption{Schematic representation of the system under investigation: An inhomogeneous 3-spin-chain is interfaced between two baths while a time-dependent field deforms the spectrum of spin 2.}
\label{fig1}
\end{figure}
The model under consideration will be a Heisenberg coupled spin-chain consisting of three spins which is a typical theoretical description for TLS. Its Hamiltonian reads
\begin{equation}
\Hop =\sum_{\mu=1}^3 \left\lbrace  \frac{\delta_\mu}{2} \op \sigma_z(\mu) + \lambda \sum_{i=x,y,z} \op \sigma_i(\mu) \otimes \op \sigma_i(\mu+1) \right\rbrace 
\label{eq1}
\end{equation}
where $\mu$ specifies the spin, $\delta_\mu$ its local energy splitting, $\lambda$ the nearest neighbour coupling strength and $\op \sigma_i(\mu)$ are the Pauli-matrices. The detuning between spin 1 and spin 3 is $\delta_{13}=(\delta_1-\delta_3)/2>0$. This model was chosen because of its analogy to the classical carnot engine. As will be explained below the spin in the middle can be considered as the "working gas" whereas the other two spins act as contacts and filters. These filters introduce a resonance effect which is essential for the model to work as a kind of carnot engine.

Thermodynamic properties can be imparted on this system by an appropriate embedding into a larger quantum environment \cite{GeMiMa2004, Henrich2005, Michel2005}, without the need of any thermal bath. In the present context, though, it is much simpler to settle for the customary approach: As depicted in Fig. \ref{fig1}, the chain is taken to be in thermal contact with two baths $\alpha=c,h$ with different  temperatures $T_c < T_h$, such that spin 1 is weakly coupled to the hot bath only and spin 3 to the cold bath. A convenient method to simulate such an open system is then based on the quantum master equation. 

To derive the master equation for our model one starts from the Liouville-von-Neumann equation (we set $\hbar=1$ and also the Boltzmann constant $k_B=1$). A typical bosonic bath model consists of uncoupled harmonic oscillators. With the help of projection operator technique up to second order and the Born-Markov approximation \cite{Kubo} the master-equation for the whole spin system then takes the form \cite{Saito2000}
\begin{equation}
\dod{}{t} \op \rho(t) =-i \kom{\Hop}{\op \rho(t)}+\op{\mathcal{D}}_h(\op \rho(t))+\op{\mathcal{D}}_c(\op \rho(t))
\label{eq2}
\end{equation}
with $\Hop$ according to (\ref{eq1}) and $\op \rho$ denoting the density matrix of the spin system. The dissipator $\op{\mathcal{D}}_\alpha$ for bath $\alpha=c,h$ is
\begin{eqnarray}
 \op{\mathcal{D}}_\alpha(\op \rho(t)) & = &(\op A_\alpha \op \Gamma_\alpha \op \rho(t))-(\op \Gamma_\alpha \op\rho(t) \op
A_\alpha) +\nonumber \\ & & +(\op \rho(t) \op \Gamma^\dagger_\alpha \op A_\alpha)-(\op A_\alpha \op \rho(t) \op
\Gamma^\dagger_\alpha)\, ,
\label{eq2b}
\end{eqnarray}
where $\op A_\alpha$ are the local coupling operators
\begin{equation}
 \op A_h=\op \sigma_x(1) \otimes \op 1(2) \otimes \op 1(3)
 \label{eq2c}
\end{equation}
\begin{equation}
 \op A_c= \op 1(1) \otimes \op 1(2) \otimes \op \sigma_x(3).
 \label{eq2d}
\end{equation}
The matrix elements of $\op \Gamma_\alpha$ in the eigenbasis have the form \cite{Henrich2005,Saito2000}
\begin{equation}
 \Bra i \op \Gamma_\alpha \Ket j=\kappa\left( 
\frac{\theta (\omega_{ij})}{\text{e}^{\omega_{ij}\beta_\alpha}-1}+\theta (\omega_{ji})\frac{\text{e}^{\omega_{ji}\beta_\alpha}}{\text{e}^{\omega_{ji}\beta_\alpha}-1}\right) A_{ij}^\alpha.
 \label{eq2e}
\end{equation}
Here $\kappa$ is a coupling parameter between system and either bath $\alpha$, $ A_{ij}^\alpha$ are the matrix elements of $\op A_\alpha$, $\theta(\omega)$ is the step function, $\beta_\alpha=1/T_\alpha$ the bath temperature and $\omega_{ij}=E_i-E_j$ are the frequencies of the total spin system ($\Hop \Ket i=E_i \Ket i$). The stationary state of this master equation with only one bath and for fixed $\delta_\mu$ is easily shown to be canonical of the form $\op \rho_\text{S}=\text{e}^{-\beta \Hop} \Tr{\{ \text{e}^{-\beta \Hop}\}}$. With both baths in place the spin chain might be viewed as a molecular bridge generating a stationary leakage current $J_L=J_h=-J_c$. 

Here the heat current $J_\alpha$ between the 3-spin-system and the bath $\alpha$ can be defined by the energy dissipated via bath $\alpha$ \cite{Breuer}:
\begin{equation}
J_\alpha = \Tr{\{ \Hop \op {\mathcal{D}}_\alpha(\op \rho(t))\} }
\label{eq3}
\end{equation}
where Tr means the trace operation. In the following a current out of bath $\alpha$ into the machine will be defined as positive. 

This situation changes drastically, if in addition, the center spin 2 is driven externally: We assume the form $\delta_2(t)=\sin(\omega t) b + b_0$ with $\omega \ll \delta_\mu$, which enables the bath to sufficiently damp spin 2. If spin 2 was driven too fast it would no longer be possible to exchange energy between spin 2 via spin 1 (3) and its respective bath. The parameters are chosen such that $\delta_2(t)$ oscillates between the two resonance conditions, i.~e. $b_0=(\delta_1+\delta_3)/2$ and $b$ is the detuning $\delta_{13}$. It should be possible to implement such devices with any TLS which allows energy exchange based on a coupling small compared with the local energy splittings. In case of an actual spin $\delta_2(t)$ could be implemented by a varying magnetic field $\vec B$. If the TLS was realized by a semiconductor quantum dot, the spectrum might be distorted by an external electric field $\vec E$.

For any diagonal state of a TLS $\mu$ in its local energy basis we can always define the temperature $T_\mu=-\delta_\mu /\left[ \ln\left(p_\mu^1/p_\mu^0\right) \right]$ and the von Neumann entropy $S_\mu=-(p_\mu^0 \ln p_\mu^0+p_\mu^1 \ln p_\mu^1)$, where $p_\mu^0$ are occupation probabilities for the ground state and $p_\mu^1$ for the excited state. Under the conditions as stated there can be no local non-equilibrium, a restrictive but convenient property of TLS. As will become transparent below, $T_2$ and $S_2$ indeed behave like thermodynamic variables.

Now we know from the Gibbs relation for the energy change of subsystem 2 that \cite{Tonner2005}
\begin{equation}
 \Delta E_2=\Delta Q_2+ \Delta W_2=\frac{1}{2} \sum_{i=0}^1 \Delta p_2^i \delta_2^i+\frac{1}{2}\sum_{i=0}^1 p_2^i\Delta \delta_2^i .
 \label{eq2f}
\end{equation}
The heat $\Delta Q_2$ is thus associated with the change of occupation probabilities $p_2^i$, the work $\Delta W_2$ with the deformation of the spectrum by external driving (modulation of $\delta_2(t)$). Over one cycle $\Delta E_2=0$ so that 
\begin{equation}
 W_2=-Q_2=-\oint T_2 \dd S_2.
 \label{eq2g}
\end{equation}
On the other hand the energy flowing from or into bath $\alpha$ is heat by definition. Over one period $\tau=2\pi/\omega$ this heat is
\begin{equation}
 Q_\alpha=\int_0^\tau J_\alpha \dd t
 \label{eq2h}
\end{equation}
allowing us to identify $Q_2=Q_c+Q_h$. The work per cycle, $W_2$, put into the machine will be counted as positive.This sign convention also applies to the transferred heat per cycle, $Q_\alpha$.

All numerical results are obtained by integrating the master equation with a 4th order Runge-Kutta algorithm. Because of $\delta_2(t)$, $\Hop$ is also time-depended. Thus for each time step one has to calculate new dissipators $\op{\mathcal{D}}_\alpha$. The parameters taken here are $\lambda = 0.01$, $\kappa=0.001$, $\delta_1=2.25$, $\delta_3=1.75$, $\omega=1/128$, $T_c=0.4$ and $T_h$ is varied, unless stated otherwise. Both coupling parameters $\lambda$ and $\kappa$ are chosen to stay in the weak coupling limit. Figure \ref{fig2} shows the dynamics of subsystem 2 in its entropy-temperature space.

In analogy with the classical Carnot machine one may approximately distinguish four different steps in the $S_2T_2$-diagram (Fig. \ref{fig2}) in combination with the respective currents in Fig. \ref{fig3}:
\begin{figure}
\centering%
\includegraphics[width=.46 \textwidth]{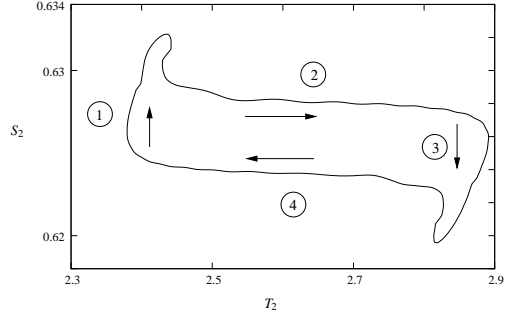}
\caption{$S_2T_2$-Diagram for the quantum heat pump. The arrows indicate the direction of circulation.}
\label{fig2}
\end{figure}

\begin{figure}
\centering%
\includegraphics[width=.46 \textwidth]{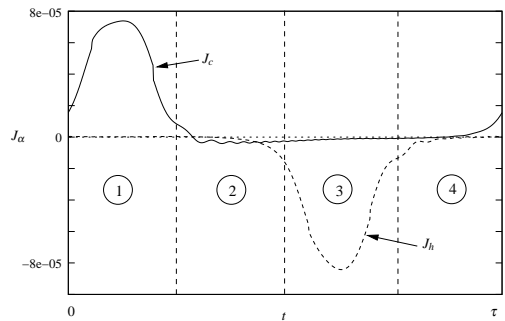}
\caption{The heat current $J_\alpha(t)$ over one cycle with duration $\tau=2 \pi/\omega=804.25$. The peaks are a result of the resonance-effect.}
\label{fig3}
\end{figure}

\begin{enumerate}
\item Quasi-iso-thermal step: Spin 2 (the "working gas") is brought in resonance with spin 3 ($\delta_2(t) \approx \delta_3$) and so couples with bath $c$ at temperature $T_c$. Because of this energy resonance the current $J_c$ via spin 3 will increase, whereas the current $J_h$ via spin 2 will be negligible. The occupation probabilities of spin 2 and 3 will approach each other and so do the respective local temperatures.
\item Quasi-adiabatic step: Spin 2 is driven out of resonance with spin 3 ($\delta_1 > \delta_2(t) > \delta_3$), now $J_c$ decrease while $J_h$ nearly stays unchanged. The occupation probability of spin 2 does not change significantly and there is almost no change in the entropy $S_2$.
\item Quasi-iso-thermal step: Spin 2 comes in resonance with spin 1 ($\delta_2(t) \approx \delta_1$) and by that in contact with bath $h$ at temperature $T_h$. $J_h$ increases whereas $J_c$ is very small. The local temperatures of spin 1 and 2 nearly equal each other.
\item Quasi-adiabatic step: Spin 2 is driven back to the starting point as in step 2.
\end{enumerate}

Also under driving some leakage current $J_L$ will flow from the hotter bath $h$ to the colder bath $c$ leading to a transferred heat $Q_L=\oint_o^\tau J_L \dd t$ per cycle. This heat is always contained in $Q_c$, $Q_h$ with the proper sign. $W_2=-(Q_h+Q_c)$ is not influenced by $Q_L$.

In contrast to the ideal Carnot heat pump which is defined in the quasi-static limit and with the bath coupling set exactly to zero during the adiabatic steps, the quantum heat pump must be driven through its cycle in a finite time (cf. the model of the endoreversible engine \cite{Curzon1975}). If one would wait after each infinitesimal time step till the system had reached its momentary steady transport configuration (i.~e. $\omega \ll \kappa$), the quantum heat pump would no longer work. The area of the $S_2T_2$-cycle would be zero. In step 1 (or 4) the local temperature of spin 2 is larger (smaller) than the respective bath temperature $T_h$ ($T_c$) which is needed to drive a current from spin 2 to bath $h$ (from bath $c$ to spin 2). There is an entropy production due the dissipation $ \op {\mathcal{D}}_\alpha$.

The efficiency of a heat pump is defined by the ratio of the heat $Q_h$ pumped to the hot reservoir and the work applied $\eta^p=Q_h/W_2$,which reduces for the Carnot heat pump to
$\eta_\text{Car}^p=T_h/(T_h-T_c)$. For the heat engine the efficiency is $\eta^e=W_2/Q_h$ which in the Carnot case leads to $\eta_\text{Car}^e=1-T_c/T_h$.

For the quantum heat pump we have calculated the work $W_2$ for one cycle out of the $S_2T_2$-diagram by (\ref{eq2g}), and the heat $Q_\alpha$ by (\ref{eq2h}). We find that indeed $W_2+Q_c+Q_h=0$ in all cases and by that confirm the use of $T_2$ and $S_2$ as effective thermodynamic variables.

\begin{figure}
\centering
 \includegraphics[width=.45 \textwidth]{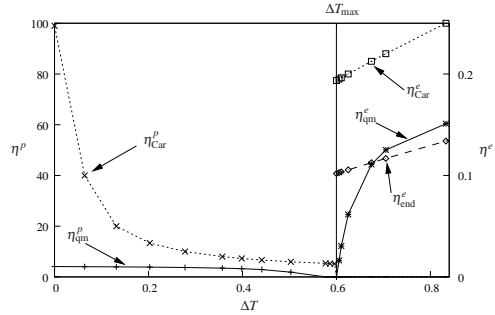}
\caption{The Carnot-efficiency $\eta_\text{Car}^p$ and the efficiency $\eta_\text{qm}^p$ of the quantum heat pump ($\Delta T < \Delta T_\text{max}$) and $\eta_\text{Car}^e$ and $\eta_\text{qm}^e$ of the heat engine ( $\Delta T > \Delta T_\text{max}$) as function of the temperature difference $\Delta T$. $\eta_\text{end}^e=1-\sqrt{T_c/T_h}$ is the efficiency of the classical endoreversible heat engine \cite{Curzon1975}.}
\label{fig5}
\end{figure}

Figure \ref{fig5} (left part) shows the numerical results for the efficiency $\eta_\text{qm}^p$ as function of  $\Delta T=T_h-T_c$ together with the corresponding Carnot efficiency $\eta_\text{Car}^p$. 
As can be seen, the efficiency of the quantum heat pump is always smaller than that of the Carnot pump, as should have been expected. Note that for $\Delta T=0$ the efficiency does not diverge as the Carnot efficiency does. Enhanced cooling extends to $\Delta T <0$, i.~e. $J_c$ can be made larger than the reversed leakage current.

\begin{figure}
\centering
 \includegraphics[width=0.4 \textwidth ]{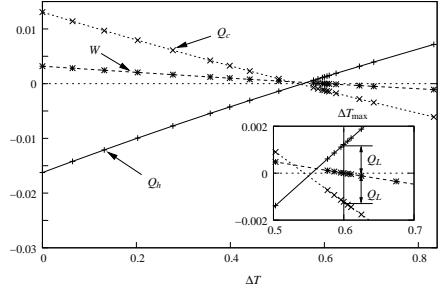}
\caption{Heat $Q_c$ and $Q_h$ as well as the work $W_2$ performed over one cycle as function of the temperature difference $\Delta T$. The inset shows these functions around the point $\Delta T=\Delta T_\text{max}$ in more detail. $Q_L$ is the leakage heat per cycle.}
\label{fig6}
\end{figure}

The work $W_2$ to be applied goes to zero at $\Delta T=\Delta T_\text{max}$ (cf. Fig. \ref{fig6}). A rough estimate gives $\Delta T_\text{max}=T_c \left(\delta_1/\delta_3-1\right) \approx 0.71$ for our model parameters. The actual numerical value is 0.60. At this point the mechanical driving becomes lossless, only the leakage remains. The efficiency $\eta_\text{qm}^p$ goes to zero at some $\Delta T < \Delta T_\text{max}$, when the heat pumped up is just cancelled by the heat leaked down, i.~e. when $Q_h=0$.

\begin{figure}
\centering
 \includegraphics[width=.4 \textwidth]{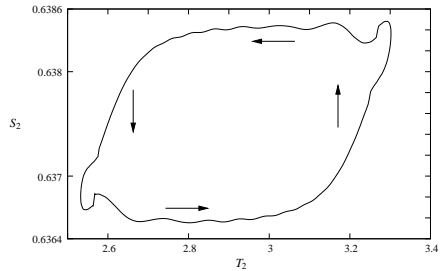}
\caption{$S_2T_2$-Diagram for the quantum heat engine. The arrows indicate the direction of circulation.}
\label{fig7}
\end{figure}

To make this conclusion clearer, also the heat $Q_c$ and $Q_h$  per cycle are included in Fig. \ref{fig6} as function of the temperature difference. The inset shows the point $\Delta T= \Delta T_\text{max}$ for which $W_2=0$ and $Q_h=Q_L=-Q_c$. $Q_c$ changes its sign before $Q_h$ and $Q_h$ before $W_2$ as $\Delta T$ approaches $\Delta T_\text{max}$ from below. For $\Delta T > \Delta T_\text{max}$ the machine starts working as a heat engine ($W_2<0$) with efficiencies $\eta_\text{qm}^e$ ($\eta_\text{Car}^e$) shown in Fig. \ref{fig5} at right. $\eta^e_{\text{qm}}$ is not far away from the classical efficiency of the so-called endoreversible heat engine, i.~e. for a machine completing its cycles at finite time \cite{Curzon1975}. Eventually this machine operates irreversibly, as the external heat flow from or into the baths is modelled as a transport process.
 Figure \ref{fig7} shows the corresponding $S_2T_2$-diagram, now executed with the opposite orientation, in agreement with the reversed sign of $W_2$.

Before concluding we remark that these 3-qubit-systems could also be used as nano-scale sensors: If $\Delta T_\text{max}$ was known as a device property, the temperature difference $\Delta T$ between an unknown bath $T_c$ and a know bath $T_h$ could be determined by varying $T_h$ until $W_2=0$. At this point the measurement was least disturbing.

Driven spin networks have long since been basis for elementary quantum information processing tasks. For these the influence of decoherence (baths) is disastrous. In this letter we have argued that,  when (two conflicting) thermal contacts are added deliberately, such networks should show a completely different, but still quite fascinating behavior. This might open up a new field of research: general quantum networks under the influence of unitary gates (interface to work reservoirs) and thermal contacts (interface to heat reservoirs). In a further twist also those reservoirs may be taken to be finite and described by quantum mechanics in full detail.




\acknowledgments
We thank H. Schmidt, H. Schr\"oder, M. Stollsteimer, J. Teifel, F. Tonner and P. Vidal for fruitful discussions. We thank the Deutsche Forschungsgemeinschaft and Landesstiftung Baden-W\"urttemberg for financial support.

\end{document}